\documentstyle[aps,prl,floats,epsf]{revtex}

\def\pdev(#1,#2;#3){\left(
           \kern-0.1em{\partial #1\over\partial #2}
           \kern-0.1em\right)_{\kern-.2em #3}}
\def\batio3{BaTiO$_3$}
\def\ave#1{\langle#1\rangle}
\def\corr(#1,#2){\ave{\Delta{#1}\Delta{#2}}}

\begin{document}

\twocolumn[\hsize\textwidth\columnwidth\hsize\csname
@twocolumnfalse\endcsname

\title{Electromechanical Behavior of \batio3\ from First Principles}

\author{Alberto Garc\'{\i}a~\cite{byline}}
\address{
Departamento de F\'{\i}sica Aplicada II, Universidad del Pa\'{\i}s
Vasco, Apdo. 644, 48080 Bilbao, Spain}

\author{David Vanderbilt~\cite{byline}}
\address{Department of Physics and Astronomy, Rutgers University, 
Piscataway, NJ 08855-0849, USA}
\maketitle

\begin{abstract}

Using an effective Hamiltonian parametrized from first-principles,
Monte Carlo simulations are performed in order to study the
piezoelectric response of \batio3\ in the ferroelectric tetragonal
phase as a function of temperature. The effect of an electric field
on the phase behavior is also illustrated by a simulation of the
transformation of a rhombohedral domain into a tetragonal one under
a strong field.

\end{abstract}

\vskip1pc]

One of the earliest technological applications of ferroelectric
materials was in the area of electromechanical transducers, the
physical basis of which is the piezoelectric effect.  The
piezoelectric coefficients characterize the (linear) change in
polarization in the presence of an external stress, or
equivalently, a change in shape under the application of an
external electric field.  While a phenomenological, descriptive
framework for such electromechanical effects has been available for a
long time, a quantitative microscopic understanding of the response of
individual materials is lacking, and further technical development has
to proceed largely by trial and error. The importance of obtaining
a deeper understanding
has been highlighted by recent reports~\cite{giant-d} of
giant piezoelectric response in single crystals of relaxor perovskites
of the form PMN-PT and PZN-PT. The structural complexity of these
materials and the variety of phenomena that might be involved make it
rather difficult to offer a clear understanding of their observed
properties.

It is in this context that first-principles calculations can
help. They can be used to selectively ``turn off'' features of the
system and study its response in conditions that are very difficult or
impossible to realize in the laboratory. Besides, they can provide a
microscopic view of the materials properties.  So far, the primary
uses of first-principles calculations relevant to this area have been
calculations of piezoelectric coefficients for simple materials at
zero temperature~\cite{zeroK-piezo}.  In this work, as a
step towards the theoretical treatment of the electromechanical
response of complex systems, we present the first calculations of
the piezoelectric response as a function of temperature for
\batio3, an important member of the perovskite family of
ferroelectric compounds.  We also illustrate the influence of
electric fields on the phase behavior of \batio3.  We base our
approach on a scheme which has proven very successful~\cite{mc-batio3}
in the description of the rich phase diagrams of perovskite oxides: an
effective Hamiltonian which contains the physically relevant degrees
of freedom of the structure (notably the ``soft mode'') is constructed
on the basis of high-quality first-principles calculations, and the
statistical mechanics of the system is then studied by Monte Carlo
simulation. The usefulness of the method is not restricted to the
calculation of piezoelectric coefficients and the study of generalized
phase diagrams.  Indeed, the elastic and dielectric response can be similarly
computed, and it can form the basis for an analysis of non-linear
effects~\cite{future}.

The appropriate thermodynamic identity for a crystal in the presence of
a (possibly anisotropic) stress and an electric field is~\cite{thermo}
\begin{equation}
dU=TdS+\sigma_\nu d\eta_\nu+E_idD_i \; \; .
\label{eq:thermo_a}
\end{equation}
Here $U$ is the internal energy of the crystal (per unit volume),
$T$ and $S$ are the temperature and entropy, $\sigma_\nu$ and
$\eta_\nu$ are the components of the stress and strain tensors in
the Voigt notation~\cite{voigt}, $E_i$ is the $i$th
component of the macroscopic electric field, and $D_i$ is the
corresponding component of the electrical displacement vector (in SI
units, ${\bf D}=\epsilon_0{\bf E} + {\bf P}$).

We parametrize the energy $U$ of the system by means of an effective
Hamiltonian $H_{\rm eff}$ which represents a Taylor expansion of the
energy surface around the high-symmetry cubic perovskite structure.
$H_{\rm eff}$ is written in terms
of the dynamical variables which are relevant to the
low-energy distortions: the amplitudes \{{\bf u}\} of the local modes
(three degrees of freedom per unit cell) which represent the ``soft''
transverse optical phonon and are directly related to the polarization
of the crystal [${\bf P}=(Z^*/V)\sum{\bf u}$, where $Z^*$ is the mode
effective charge and $V$ is the cell volume];
a set \{{\bf v}\} of displacement variables
representing the acoustic modes; and the six components of the
homogeneous strain $\eta$. The parameters of the energy expansion,
including the on-site local-mode self-energy, the interaction
between local modes (both short-range and dipole-dipole), the elastic
energy, and the local mode-elastic coupling, are computed using highly
accurate first-principles LDA calculations with Vanderbilt ultrasoft
pseudopotentials~\cite{uspp}. 
More details about the construction of the effective
Hamiltonian are given in Ref.~\onlinecite{mc-batio3}, where the method
is shown to provide a good account of the phase transition sequence in
\batio3\ within the limitations of an approach based on low-energy
distortions. The extension
of the standard Metropolis Monte Carlo algorithm to include the
effects of stress and electric field involves replacing the Boltzmann
probability factor $\exp(-\beta U^j)$ by
$\exp[-\beta(U^j-\sigma_\nu\eta^j_\nu-E_iP^j_i)]$ in the acceptance
criterion for state $j$. For a given temperature, stress, and field,
the strain $\eta$ and the mode variables are allowed to fluctuate,
their average values determining the shape and net polarization of the
system.

As an important application of the method, we compute the piezoelectric
response of the tetragonal (ferroelectric) phase of \batio3\ (point
group $4mm$), which is stable from approximately 278K to 403K and
exhibits a spontaneous polarization that we take to be
along the $z$ axis. The relevant
coefficients are
\begin{equation}
d_{i\nu}
=\pdev(\eta_\nu,E_i;\sigma)
=\pdev(P_i,\sigma_\nu;E)
=\beta\,\corr(P_i,\eta_\nu)
\end{equation}
where $\Delta X=X-\ave{X}$ and the averages are computed using the
extended Boltzmann factor defined above~\cite{explain-corr}. These
equations suggest three different ways to calculate the response:
direct calculations of the average strain as a function of applied
field, or the average polarization as a function of (anisotropic)
stress, and computation of the statistical correlation between
polarization and strain. The latter is conceptually the simplest,
although relatively long simulations (on the order of 100000 Monte
Carlo sweeps (MCS)~\cite{mcs}) are needed to obtain good
statistics. Of the direct approaches, the field-strain calculations
are the most efficient as only one series of calculations for varying
$E_3$ is needed to compute the most common coefficients $d_{31}$ and
$d_{33}$ (representing respectively the transverse contraction and the
longitudinal expansion under the application of a field parallel to
the ferroelectric axis).

\begin{figure} 
\epsfxsize=3.3 truein
\epsfbox{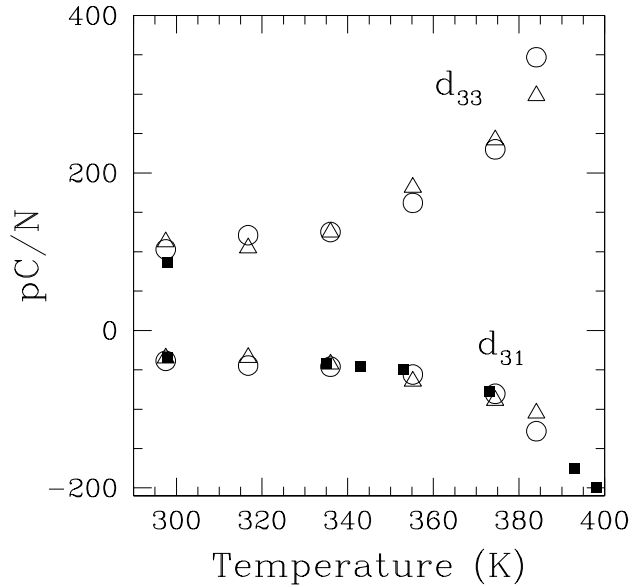}
\caption{Piezoelectric coefficients as a function of temperature in
the tetragonal phase. Open triangles and circles represent values computed
from the strain response to an electric field and from
polarization-strain correlations, respectively. Solid squares show
experimental data points (Ref.~\protect\onlinecite{d-exp}).}
\label{fig:tet.piezo} 
\end{figure}

An effective Hamiltonian based on a finite Taylor expansion of the
energy should not be expected to reproduce perfectly the behavior of
the material at relatively high temperatures. In particular (see
Ref.~\onlinecite{mc-batio3}) the theoretical transition temperatures
are progressively shifted downwards with respect to the true
ones~\cite{tcs}. In order to provide a better comparison of our
results for $d_{31}$ and $d_{33}$ to experiment
(Fig.~\ref{fig:tet.piezo}), we have therefore rescaled linearly the
theoretical temperatures so that the end points of the range of
stability of the tetragonal phase coincide. The agreement with the
available experimental data is very good, and the general trend of the
temperature dependence (upward tendency in $d_{33}$, downward in
$d_{31}$) corresponds to the pseudo-divergent behavior observed
experimentally near the tetragonal-to-cubic
transition~\cite{exp-divergent}.
In the thermodynamic limit, the correlation and field-strain
approaches should be completely equivalent. Since we use finite runs
and finite simulation boxes ($12\times 12\times 12$ or 1728 unit cells
for the correlation analysis and $10\times 10\times 10$ or 1000 unit
cells for the field-strain calculations), the results are similar
but not identical.

\begin{figure} 
\epsfxsize=3.3 truein
\epsfbox{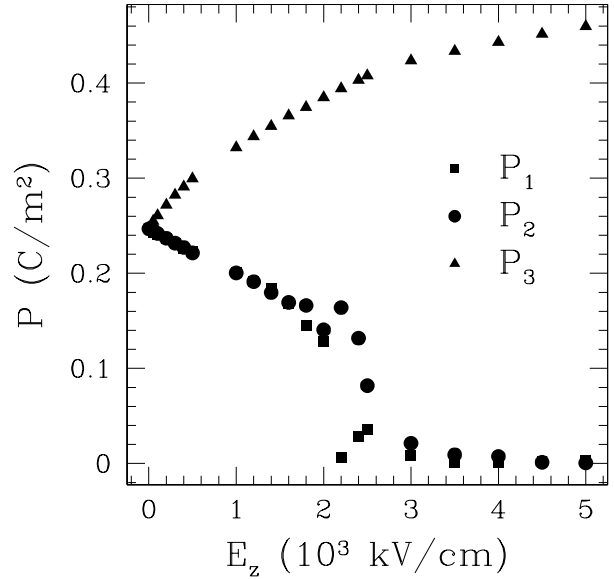}
\caption{Polarization vs.\ field for the phase transformation from
rhombohedral to tetragonal under a strong field.}
\label{fig:r2t.p} 
\end{figure}

Our method can also be applied to the study of the stability of the
different phases as a function of the external field and
stress. As an illustration of the influence of electric fields on the
phase diagram, we performed a simulation of the effect of a strong
field on the strain and electrical polarization of a \batio3\ sample
in the rhombohedral (R) phase. This phase, with point group $3m$, is
stable at temperatures below 183K, and exhibits a spontaneous
polarization along one of the original $\langle111\rangle$
cubic directions, and a
corresponding strain deformation with respect to the cubic phase satisfying
$\eta_1=\eta_2=\eta_3$ and $\eta_4=\eta_5=\eta_6$. Choosing a
simulation temperature of 100K, we applied an electric field along the
cubic $z$ axis, progressively increasing the strength of the field
up to a maximum value of $5\times 10^3$ kV/cm.
\begin{figure} 
\epsfxsize=3.3 truein
\epsfbox{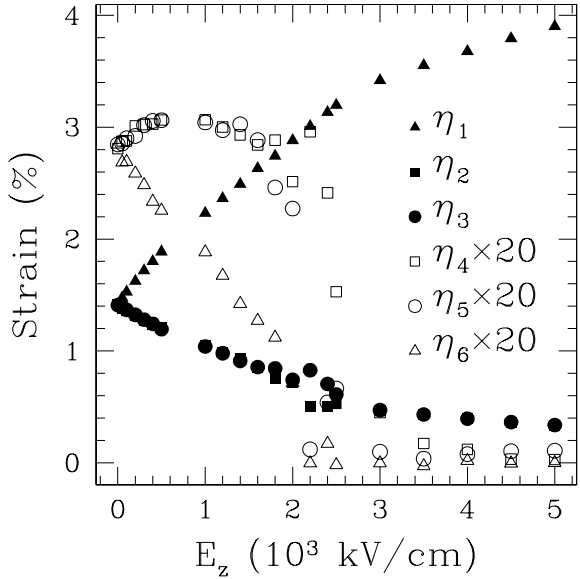}
\caption{Strain vs.\ field for the phase transformation from
rhombohedral to tetragonal under a strong field. Off-diagonal strain
components are magnified for clarity.}
\label{fig:r2t.strain} 
\end{figure} 
The evolution of the strain and polarization as a function of the
field is shown in Figs.~\ref{fig:r2t.strain} and~\ref{fig:r2t.p},
respectively. The rhombohedral symmetry is immediately broken by the
field, as can be seen by the $\eta_1=\eta_2\neq\eta_3$ splitting (and
the analogous one among the polarization components). In the
approximate range $2-2.5\times 10^3$ kV/cm, there is a noticeable
split in $\eta_1$ and $\eta_2$, all the off-diagonal components of the
strain save $\eta_4$ ($2\eta_{23}$ in the standard tensor notation)
fall to zero, and $P_1$ falls to nearly zero. Thus, this range
corresponds to an orthorhombic phase with the polarization in the $yz$
plane (different simulations ``pick'' at random
between the $yz$ and the $xz$ planes). For larger fields, the
equality between $\eta_1$ and $\eta_2$ is restored, all the
off-diagonal strain components fall to zero, and $P_2$ follows $P_1$
in its earlier drop. The resulting phase is tetragonal, with {\bf P}
along the [001] direction.
The behavior emerging from our simulation parallels that seen
experimentally~\cite{giant-d} in the PMN-PT and PZN-PT
giant-response materials: individual domains are in a rhombohedral
structure, and as an electric field is applied and progressively
increased along $z$, the polarization rotates from a [111] direction
to a [001] direction.

In conclusion, we have shown how the electromechanical response of a
system can be computed from first principles using an effective
Hamiltonian suitably augmented by terms representing the influence of
applied (anisotropic) stresses and electric fields. As an application,
we have presented the first calculations of the piezoelectric response
of ferroelectric tetragonal \batio3\ as a function of temperature. We
have also performed simulations illustrating the effect of electric
fields on the phase behavior of this material. 

This work was supported in part by the ONR Grant N00014-97-1-0048
and by the UPV research grant 060.310-EA149/95. We thank J.M. Perez-Mato and 
Karin Rabe for useful comments.


\begin{references}
\bibitem{explain-corr} The relationship between $d_{i\nu}$ and the
strain and polarization correlations can be simply obtained
by differentiation of the expression $\Lambda\ave{X}=\sum_j{X^j
{\rm Prob}_j}$, where Prob$_j$ is the extended Boltzmann factor and
$\Lambda$ is the extended partition function.
\bibitem{mcs} A Monte Carlo sweep is completed after each local
variable is considered for a ``flip attempt'' and each component of
the homogeneous strain suffers $2L+1$ attempted changes, where $L$ is
the linear size of the simulation box.
\bibitem{tcs} For the rhombohedral to orthorhombic to tetragonal
to cubic transition sequence, the experimental $T_c$'s are
respectively 187K, 278K, and 403K, while the theoretical ones
are 197K, 230K, and 295K.
\end{references}
\end{document}